\documentclass[manuscript]{aastex}
\newcommand{\be}{\begin{equation}}
\newcommand{\ee}{\end{equation}}
\newcommand{\bary}{\begin{eqnarray}}
\newcommand{\eary}{\end{eqnarray}}

\shorttitle{How many  Ultra High Energy Cosmic Rays could we expect from Centaurus A?}
\shortauthors{Fraija, Perez, Gonzalez and Marinelli}

\begin{document}
\title{How many  Ultra High Energy Cosmic Rays could we expect from Centaurus A?}
\author{N. Fraija\altaffilmark{1}, M. M. Gonz\'alez\altaffilmark{1},  M. Perez\altaffilmark{1},  A. Marinelli\altaffilmark{2}}
\affil{Instituto de Astronom\'ia, UNAM, M\'exico, 04510}
\affil{Instituto de F\' isica, UNAM, M\'exico, 04510}
\email{nifraija@astro.unam.mx,  magda@astro.unam.mx,  jguillen@astro.unam.mx, antonio.marinelli@fisica.unam.mx}
\altaffiltext{1}{Instituto de Astronom\' ia, Universidad Nacional Aut\'onoma de M\'exico, Circuito Exterior, C.U., A. Postal 70-264, 04510 M\'exico D.F., M\'exico}
\altaffiltext{2}{Instituto de F\' isica, Universidad Nacional Aut\'onoma de M\'exico, Circuito Exterior, C.U., A. Postal 70-264, 04510 M\'exico D.F., M\'exico}


\begin{abstract}

The Pierre Auger Observatory has associated a few ultra high energy cosmic rays  with the direction of Centaurus A.
This source  has been deeply studied in radio, infrared, X-ray and $\gamma$-rays (MeV-TeV) because it is the nearest radio-loud active galactic nuclei.
 Its spectral energy distribution or spectrum shows two main peaks, the low energy peak, at an energy of  $10^{-2}$ eV,  and the high energy peak, at about $150$ keV.
 There is also a faint very high energy (E $\geq$ 100 GeV) $\gamma$-ray emission fully detected by the High Energy Stereoscopic System experiment.  In this work  we describe the entire spectrum,
the two main peaks with a Synchrotron/Self-Synchrotron Compton model and, the Very High Energy emission with a hadronic model.  We consider p$\gamma$ and $pp$ interactions. For the p$\gamma$ interaction, we  assume that the target photons  are those produced at  150 keV in the leptonic processes. On the other hand, for the pp interaction we consider as targets the thermal particle densities in the  lobes. Requiring a satisfactory description of the spectra 
at very high energies with p$\gamma$ interaction we obtain  an excessive luminosity in ultra high energy cosmic rays (even exceeding  the Eddington luminosity). However, when considering  pp interaction to describe the $\gamma$-spectrum, the obtained number of ultra high energy cosmic rays are in agreement with Pierre Auger observations.  Moreover,  we calculate  the possible neutrino signal from  pp  interactions on a Km$^3 $ neutrino telescope using Monte Carlo simulations.


\end{abstract}



\section{Introduction}

Centaurus A (Cen A) is a Fanaroff $\&$ Riley Class I (FRI) active galactic nuclei (AGN). At a distance of 3.8 Mpc, it is the nearest radio-loud AGN and an excellent source for studying the physics of relativistic outflows and radio lobes.  Its giant radio lobes, which subtend $\sim\,10^\circ$ on the sky, are oriented primarily in the north-south direction. They were imaged at 4.8 GHz by the Parkes  telescope \citep{jun93} and studied up to $\sim\,60$ GHz by  Hardcastle et al. 2009 utilizing the Wilkinson Microwave Anisotropy Probe  \citep[WMAP;][]{hin09}.   Cen A has a  jet with an axis subtending an angle to the line of sight estimated as $15^\circ\,-\,80^\circ$ \citep[see, e.g.] [and reference therein]{hor06}.   Cen A has been well studied in radio, infrared, optical \citep{ win75,mus76,bow70,bai81},  X-ray and $\gamma$-rays (MeV-TeV) \citep{har03,sre99,aha09}. 
A tentative detection (4.5 $\sigma$) of Cen A  at very high energy (VHE)  in the 1970s was reported by  Grindlay et al. 1975. Subsequent VHE observations made with Mark III \citep{car90}, JANZOS \citep{all93} and CANGAROO \citep{row99, kab07}  experiments resulted in flux upper limits.  Cen A was also detected from  MeV to GeV energies by all instruments on board of the Compton Gamma-Ray Observatory (CGRO) in the period of 1991-1995, revealing a peak in the spectral energy distribution (SED) in $\nu F_\nu$ representation at $\sim0.1$ MeV with a maximum flux of about $\sim10^{-9}$ erg cm$^{-2}$ s$^{-1}$ \citep{ste98}.  For more than 120 hr Cen A was observed \citep{aha05, aha09} by  High Energy Stereoscopic System (H.E.S.S.) experiment.  A signal with a statistical significance of $5.0\sigma$ was detected from the region including the radio core and the inner kpc jets. The integral flux  above an energy threshold of $\sim250$ GeV was measured to be $0.8\%$ of the Crab Nebula  (apparent luminosity: L($>$250 GeV) $\sim 2.6\times 10^{39}$ erg s$^{-1}$). The spectrum was described by a power law with a spectral index of $2.7\pm0.5_{stat}\pm0.2_{sys}$.  No significant flux variability was detected in the data set. Also, for a period of 10 months, Cen A was monitored by Large Area Telescope (LAT) on board the Fermi Gamma-Ray Space Telescope.  Flux  levels  were not significantly different from those found by the Energetic Gamma Ray Experiment Telescope (EGRET). However, the LAT spectrum was  described with a photon index  of $2.67\pm0.10_{stat}\pm0.08_{sys}$ \citep{abd10}.    The spectra recorded by the cited gamma-ray experiments can be considered as the intrinsic spectra of Cen A because they are not affected by the Extragalactic Background Light (EBL) absorption.

It has been proposed that astrophysical sources accelerating ultra high energy cosmic rays (UHECRs) also could produce
high energy $\gamma$-rays by proton interactions with photons at the source and/or the surrounding radiation and matter.  Hence, VHE photons detected from Cen A could be the result of hadronic interactions of cosmic rays accelerated by the jet with photons radiated inside the jet 
or protons in the lobes \citep{gop10, rie09, kac09a, kac09b, rom96, iso02, hon09, abd10, der09}.

Pierre Auger Observatory (PAO) studied the spectra of UHECR above $57$ EeV through their shower properties finding a mixed composition of $p$ and $Fe$ \citep{yam07, abr08, ung07}.  By contrast,  HiRes data are consistent with a dominant proton composition at those energies, but uncertainties in the shower properties  \citep{ung07} and in the particle physics extrapolated to this extreme energy scale \citep{eng07} preclude definitive statements about the composition. At least two events of the UHECRs observed by PAO were detected \citep{abr07,abr08} inside of a $3.1^{\circ}$ circle centered at Cen A.

Synchrotron/Synchrotron-Self Compton (SSC) models have been very successful in explaining the multiwavelength emission from Broad-Line Lacertae (BL Lac) objects \citep{blo96, tav98}. If FRIs are misaligned BL Lac objects, then one would  expect synchrotron/SSC to explain their non-thermal spectral energy distribution (SED) as well. In the synchrotron/SSC scenario the low energy  emission, radio through optical, originates from synchrotron radiation while high energy emission, X-rays through VHE $\gamma$-rays, originates from SSC.  
However, many blazars have higher energy synchrotron peaks,  so this mechanism then covers much of the X-ray band; for them only the $\gamma$-rays come from SSC mechanism. In Cen A, synchrotron/SSC model has been applied successfully to fit the two main peaks of the SED, jointly or separated, with one or more electron populations \citep{abd10, chi01, len08, ore09, har09}.   On the other hand, some authors \citep{der09, gup08, bec09} have considered  hadronic processes to explain the VHE photons apparent in the SED.

In this work we use the fact  that leptonic processes are insufficient to explain the entire spectrum of Cen A, and introduce hadronic processes that may
leave a signature in the number of UHECRs observed on Earth.  Our contribution is to describe jointly the SED of Cen A as well as the observed number of UHECR by PAO. We first require a description of the SED up to the highest energies obtaining parameters as:  proton spectral index ($\alpha_p$),  proton proportionality constant ($A_p$) and  the normalization energy ($E_0$). Then, we use these parameters to estimate the expected UHECRs observed by PAO. The main assumption here, is the continuation of the proton spectrum to ultra high energies. We also estimate the neutrino expectation in a hypothetical Km$^3$ telescope when considering that the VHE photons in the SED of Cen A  are produced by pp interaction. 


\section{UHECRs from Cen A}

The Pierre Auger Observatory,  localized in the Mendoza Province of Argentina at $\approx$ 36$^\circ$ S latitude,  determines the arrival directions and energies of UHECRs using four fluorescence telescope arrays and 1600 surface detectors spaced 1.5 km.  The large exposure of its ground array, combined with accurate energy and arrival direction measurements, calibrated and verified from the hybrid operation with the fluorescence detectors, provides an opportunity to explore the spatial correlation between cosmic rays and their sources in the sky.
The Pierre Auger Collaboration reported an anisotropy in the arrival direction of UHECRs  \citep{abr07,abr08}. While a possible correlation with nearby  AGNs is still under discussion, it has been pointed out that some of the events can possibly be associated with Cen A \cite[e.g.][]{gor08,mos09,kac09a}.

The corrected PAO  exposure for a point source is given by $\Xi\,t_{op}\, \omega(\delta_s)/\Omega_{60}$, where $\Xi\,t_{op}=(\frac{15}{4})\,9\times10^3\,\rm km^2\,yr$, $t_{op} $ is the total operational time (from 1st January 2004 until August 31st, 2007),  $\omega(\delta_s)\simeq 0.64$ is an exposure correction factor for the declination of Cen A, and $\Omega_{60}\simeq\pi$ is the Auger acceptance solid angle \citep{cuo09}. For a proton power law with spectral index $\alpha_p$ and proportionality constant $A_p$, the expected number of UHECRs from Cen A observed by PAO above an energy, $\rm E_{min}$, is given by,

\bary
N_{\tiny UHECR}=\frac{\Xi\,t_{op}\, \omega(\delta_s)}{(\alpha-1)\,\Omega_{60}}\,A_p\,E_0\,\left(\frac{E_{min}}{E_0}\right)^{-\alpha_p+1}
\eary

\noindent where $E_0$ is the normalization energy. In other words, the expected number of UHECRs  depends on the proton spectrum parameters. If 
we assume that protons at lower energies have hadronic interactions responsible for producing the observed gamma-ray spectra at very high
energies then we can estimate these parameters. An interesting quantity is the apparent isotropic UHECR luminosity that also depends on the spectrum
parameters as,

\bary
L_p= \frac{4\,\pi\, d_z^2\,A_p\,E_0^2}{(\alpha_p-2)} \,\left(\frac{E_{min}}{E_0}\right)^{2-\alpha_p}
\eary

\noindent where $ d_z$ is the distance to Cen A.  On the other hand,  during flaring intervals the apparent isotropic jet power can reach $\approx 10^{46}$erg s$^{-1}$, hence the maximum particle energies of a cold relativistic wind with velocity $\beta$, apparent isotropic luminosity (L),  Lorentz factor ($\Gamma$) and  equipartition parameter of the  magnetic field ($\epsilon_B$) is given by \citep{der09},

\begin{equation}
E_{max}\approx3\times10^{20}\frac{\sqrt{\epsilon_B\,L/10^{46}\, erg\, s^{-1}}}{\beta^{3/2}\,\Gamma}\,eV
\end{equation}

\noindent where  $\Gamma=1/\sqrt{1-\beta^2}$.

\section{Leptonic Synchrotron/SSC Model}

In accordance to the AGN model presented in detail by Becker $\&$ Biermann 2009,  the  synchrotron photons  come from internal shocks in the jet. The associated  electrons are accelerated by the first Fermi mechanism \citep{bla76} and the non-thermal electron spectrum can be described by broken power-law given by,

\begin{equation}
\label{espele}
\frac{dN_e}{dE_e}   = A_e
\cases {
(\frac{E_e}{E_0})^{-\alpha} 						& 	$E_{e,m}<E_e < E_{e,c}$\cr
(\frac{E_{e,c}}{E_0}) (\frac{E_e}{E_0})^{-(\alpha+1)}          & 	$E_{e,c} \leq  E_e<E_{e,Max}$\cr
}
\end{equation}

\noindent where  $A_e$ is the proportionality electron constant, $\alpha$ is the electron spectral index, $E_{e,i}=\gamma_{e,i} m_e c^2$  and $\gamma_{e,i}$ is the electron Lorentz  factor. The index $i$ is m, c or Max for minimum, break and Maximum, respectively.  For instance, $\gamma_{e,m}$  is the minimum electron Lorentz factor.  $\gamma_{e,i}$ is given \citep{vie95,gal99,che96} as follows,

\bary\label{lorentz}
\gamma_{e,m}&=&1836.15\,\frac{(\alpha-2)}{(\alpha-1)}\,\epsilon_e\,\Gamma\cr
\gamma_{e,c}&=&548. 48 (1+z)^{-3}\,   \frac{ f_{es}\,\Gamma^2\, \delta^3_D\,\epsilon^{-1}_B }{ \beta^2}\, \biggl( \frac{ L^{obs} } {5 \times10^{43} erg s^{-1} } \biggr)^{-1}\,\biggl(\frac{dt^{obs}}{2.6 \times10^6 s}\biggr)   \cr
\gamma_{e,Max}&=&1.11\times 10^8  \,(1+z)^{-1}\,\frac{\delta_D\,\Gamma^{1/2}\,\epsilon_B^{-1/4}}{\beta}\,\biggl( \frac{ L^{obs} } {5 \times10^{43} erg s^{-1} } \biggr)^{-1/4}\,\biggl(\frac{dt^{obs}}{2.6 \times10^6 s}\biggr)^{1/2}
\eary 

\noindent  where $\epsilon_e$ is  the electron energy shock, $\delta_D\equiv [\Gamma(1-\beta\mu)]^{-1}$ is the  Doppler factor,   $\mu=\cos\theta$ is the observing angle  along the line of sight, $L^{obs}$ is the observed luminosity, $dt^{obs}$ is the variability and $f_{es}$ is the ratio of shell expansion time to synchrotron emission time given by \citep{bat03},

\be
f_{es}(E_e)=
 \left\{
\begin{array}{l l}
 \frac{E_e}{E_{e,c} } & \quad  E_e < E_{e, c} \\
  1 & \quad E_e \ge E_{e,c}\\
\end{array} \right. .
\ee.

The magnetic field,  which comes from an equipartition law,  is given by

\be
B=2.6\, \rm G  \, \epsilon^{1/2}_B\, \delta_D^{-2}\,\Gamma^{-1}\,\biggl( \frac{ L^{obs} } {5 \times10^{43} erg s^{-1} } \biggr)^{-3/2}\,\biggl(\frac{dt^{obs}}{2.6 \times10^6 s}\biggr)
\ee

\noindent As the  electrons are accelerated in the shock inside the magnetic field  B, they emit photons by synchrotron radiation depending on  the electron Lorentz factor. The photon energy in the source frame is related to the photon energy in the Earth's frame by $E^{obs}_{\gamma} = \frac{\delta_{D}}{1+z} E_{\gamma}$\citep{cha09b} then, the  observed energies \citep{ryb79} using equation \ref{lorentz} are given by,

\begin{eqnarray}\label{synrad}
E^{obs}_{\gamma,m,syn} &=&0.12\,\rm {eV}\,\frac{(\alpha-2)^2}{(\alpha-1)^2}(1+z)^{-1}\,\delta^{-1}_D\,\Gamma\, \epsilon_{e}^2\, \epsilon_B^{1/2}\, \biggl( \frac{ L^{obs} } {5 \times10^{43}\, \rm{erg\, s^{-1}} } \biggr)^{1/2}\,\biggl(\frac{dt^{obs}}{2.6 \times10^6 s}\biggr)^{-1}\cr
E^{obs}_{\gamma,c,syn} &=&0.01\,\rm{eV}\, (1+z)^{-5}\, \frac{f_{es}^2\,\delta_D^5\, \Gamma^3\, \epsilon_B^{-3/2}}{\beta^4}\,\biggl( \frac{ L^{obs} } {5 \times10^{43}\,\rm{ erg\, s^{-1}} } \biggr)^{-3/2}\,\biggl(\frac{dt^{obs}}{2.6 \times10^6 s}\biggr)\cr
E^{obs}_{\gamma,Max,syn} &=& 4.3\times 10^8\,\rm{eV}\,(1+z)^{-1}\, \frac{\delta_D}{\beta^2}
\end{eqnarray}

\begin{figure}
\plotone{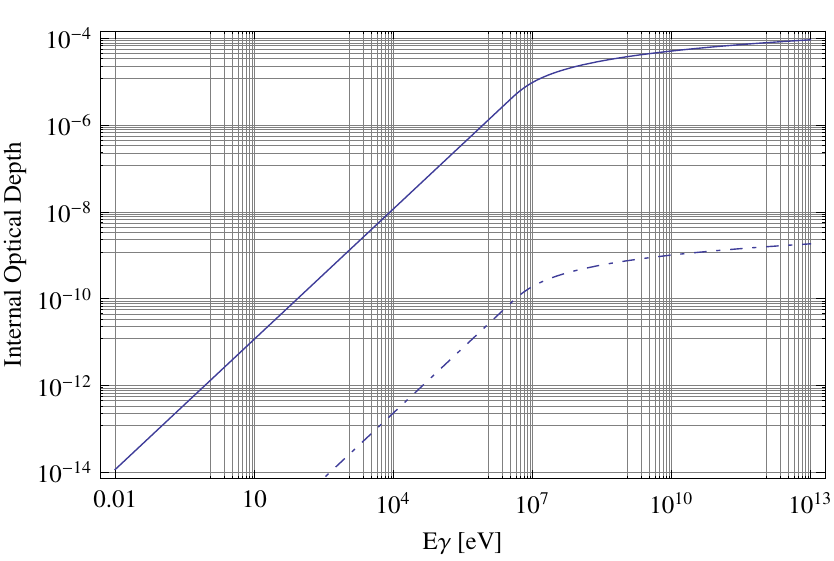}
  \caption{Internal optical depth $\tau_{\gamma\gamma}$ as a function of photon energy $E_\gamma$ for two different luminosities: $L^{obs}=5.0\times 10^{43}$erg/s (solid line) and $L^{obs}=1.0\times 10^{39}$erg/s (dot-dashed line)}
  \label{optdep}
\end{figure}

We notice that the observed energies correspond to the cut-off  energies in the synchrotron range  and depend on parameters, as $f_{es}$ and $\delta_D$, that can be determined by fitting the first peak of the SED. On the other hand, the differential spectrum, $dN_\gamma/dE_\gamma$, of the synchrotron photons is related to the electron spectrum through  $ f_{es}(E_e)\,E_e\,(dN_e/dE_e)\,dE_e=E_\gamma\,(dN_\gamma/dE_\gamma)\,dE_\gamma$, where  $E_\gamma= C_e\, E^2_e $ and $C_e$ is given in eq. (\ref{Ce}).  Thus,  we can obtain the observed synchrotron spectrum as follow \citep{gup07}

\begin{equation}
\label{espsyn}
\biggl(E^2_\gamma\, \frac{dN_\gamma}{dE_\gamma}\biggr)^{obs}_{syn} = A_{e,\gamma}
\cases {
(\frac{E_{\gamma,m,syn}}{E_{0}})^{-4/3 -(\alpha-3)} (\frac{E_{\gamma,c,syn}}{E_{0}})^{-1/2} (\frac{E_{\gamma,syn}}{E_{0}})^{4/3}    &  $E^{obs}_{\gamma,syn} < E^{obs}_{\gamma, m,syn}$\cr
(\frac{E_{\gamma,c,syn}}{E_{0}})^{-1/2} (\frac{E_{\gamma,syn}}{E_{0}})^{-(\alpha-3)/2}                                                                                 &  $E^{obs}_{\gamma, m,syn} < E^{obs}_{\gamma,syn} < E_{\gamma, c,syn}$\cr
(\frac{E_{\gamma,syn}}{E_{0}})^{-(\alpha-2)/2}                                                                                                                                                           &  $E^{obs}_{\gamma, c,syn} < E^{obs}_{\gamma,syn} < E^{obs}_{\gamma,Max,syn} $\cr
}
\end{equation}

\noindent where
\be
 A_{e,\gamma}=2.16\times 10^{-17}\,\frac{\Gamma^2\, \delta^2_D\, E_0^2\, A_{e} \,e^{-\tau_{\gamma\gamma} }}{(1+z)^2}\,\biggl(\frac{dt^{obs}}{2.6 \times10^6 s}\biggr)^2\,\biggl(\frac{d_z}{3.8\,\rm{Mpc}}\biggr)^{-2} (C_e E_0)^{\frac{\alpha-2}{2}}
\ee
\be\label{Ce}
C_e=1.35\times 10^{-19}\,(1+z)^2\,\epsilon_B^{1/2}\,\delta_D^{-2}\,\Gamma^{-1}\, \biggl( \frac{ L^{obs} } {5 \times10^{43}\,\rm{erg\, s^{-1}} } \biggr)^{1/2}\,\biggl(\frac{dt^{obs}}{2.6 \times10^6 s}\biggr)^{-1}\,eV^{-1}
\ee

\noindent and $\tau_{\gamma\gamma}$ is the optical depth (see fig. \ref{optdep}).  Equation \ref{espsyn} represents the low energy contribution (IR  to optical emission ) to the whole spectrum.

\noindent To obtain the  contribution from X-ray to $\gamma$-rays,  we assume that  the relativistic electrons inside the jet  can upscatter the synchrotron photons  in the same knot up to  higher energies in accordance with,

\begin{equation}
E^{obs}_{\gamma,m,SSC}\sim \gamma^2_{e,m} \,E^{obs}_{\gamma,m,syn}, \hspace{1cm} E^{obs}_{\gamma,c,SSC} \sim \gamma^2_{e,c} \, E^{obs}_{\gamma,c,syn},  \hspace{1cm} E^{obs}_{\gamma,Max,SSC} \sim  \gamma^2_{e, Max}\,  E^{obs}_{\gamma,Max,syn}
\end{equation}

Now, with eqs. (\ref{lorentz}) and (\ref{synrad}) we can finally obtain the inverse Compton photon energies,

\begin{eqnarray}
E^{obs}_{\gamma,m,SSC} &=& 4.01\times 10^5\,\rm{eV}  \frac{(\alpha-2)^4}{(\alpha-1)^4} (1+z)^{-1}\, \epsilon_e^4\, \epsilon_B^{1/2}\, \Gamma^3\, \delta_D^{-1}\, \biggl( \frac{ L^{obs} } {5 \times10^{43}\,\rm{erg\, s^{-1}}} \biggr)^{1/2}\,\biggl(\frac{dt^{obs}}{2.6 \times10^6 s}\biggr)^{-1}   \cr
E^{obs}_{\gamma,c,SSC} &=& 3.19\times 10^3\,\rm{eV}\, (1+z)^{-11}\,(1+x)^{-4} \, \frac{f^4_{es}\,\epsilon_e^{-7/2}\, \delta_D^{11}\, \Gamma^7}{\beta^3}\,  \biggl( \frac{ L^{obs} } {5 \times10^{43}\,\rm{erg\, s^{-1}} } \biggr)^{-7/2}\,\biggl(\frac{dt^{obs}}{2.6 \times10^6 s}\biggr)^{3}     \cr
E^{obs}_{\gamma,max,SSC} &=& 5.4\times 10^{24}\,\rm{eV}\, (1+z)^{-3}\, \frac{\epsilon_B^{-1/2}\, \delta_D^3\, \Gamma}{\beta^4}\, \biggl( \frac{ L^{obs} } {5 \times10^{43}\,\rm{erg\, s^{-1}} } \biggr)^{-1/2}\,\biggl(\frac{dt^{obs}}{2.6 \times10^6 s}\biggr)
\end{eqnarray}

The Self Synchrotron Compton Spectrum is generally \citep{fra04} written as,

\begin{equation}\label{Ae}
A_{ic}\,\biggl(\frac{dN_\gamma}{dE_\gamma}\biggr)_{SSC}=\frac{1}{E_{\gamma,IC}} \int \frac{dN_e}{dE_e}\,dE_e\, \int \biggl(\frac{dN_\gamma}{dE_\gamma}\biggr)_{syn}\,dE_\gamma
\end{equation}

Combining Eqs. (\ref{espsyn}) and (\ref{espele}) with (\ref{Ae}), we have that the observed Self-Synchrotron Compton  spectrum is given by

\begin{equation}
\label{espic}
\biggl(E^2_\gamma\, \frac{dN_{\gamma}}{dE_{\gamma}}\biggr)^{obs}_{SSC} \simeq A_{\gamma,SSC}
\cases{
(\frac{E_{\gamma,m,SSC}}{E_{0}})^{-4/3 -(\alpha-3)} (\frac{E_{\gamma,c,SSC}}{E_{0}})^{-1/2} (\frac{E_{\gamma,SSC}}{E_{0}})^{4/3}            & $E^{obs}_{\gamma,SSC} < E^{obs}_{\gamma,m,SSC}$\cr
(\frac{E_{\gamma,c,SSC}}{E_{0}})^{-1/2} (\frac{E_{\gamma,SSC}}{E_{0}})^{-(\alpha-3)/2}                                                                                       &  $E^{obs}_{\gamma,m,SSC} < E^{obs}_{\gamma,SSC} < E_{\gamma,c,SSC}$\cr
(\frac{E_{\gamma,SSC}}{E_{0}})^{-(\alpha-2)/2}                                                                                                                                                               &  $E^{obs}_{\gamma,c,SSC} < E^{obs}_{\gamma,SSC}< E^{obs}_{\gamma,max,SSC}$\cr
}
\end{equation}

\noindent where,

\begin{eqnarray}
 A_{\gamma,SSC} &=&2.32\times 10^{-19} \frac{\Gamma^2\, \delta^2_D\, A_{ic}^{-1} A_e^2\,E_0 \,e^{-\tau_{\gamma\gamma} }}{(1+z)^2}\biggl(\frac{dt^{obs}}{2.6 \times10^6 s}\biggr)^2\,\biggl(\frac{d_z}{3.8\,\rm{Mpc}}\biggr)^{-2}\, (C_e\,E_0)^{\frac{\alpha-2}{2} }\cr
 && (\frac{0.505\,MeV}{E_{0}})^{-(\alpha - 1)} \biggr(\frac{E_{\gamma,c,syn}}{1\,\rm{eV}}\biggr)^{-1/2} \biggl(\frac{E_{\gamma,ic}}{150\,\rm{keV}}\biggr)^{-1/2}
\end{eqnarray}

\noindent Summarizing, the Leptonic model describes the whole spectrum at energies below a few tens of GeV as stated by equations \ref{espsyn} and \ref{espic} considering,

\begin{equation}
\biggl(E^2_\gamma\, \frac{dN_{\gamma}}{dE_{\gamma}}\biggr)^{obs}_{Lept\,model}=\biggl(E^2_\gamma\, \frac{dN_{\gamma}}{dE_{\gamma}}\biggr)^{obs}_{syn}+\biggl(E^2_\gamma\, \frac{dN_{\gamma}}{dE_{\gamma}}\biggr)^{obs}_{SSC}
\end{equation}

\section{Hadronic Model}

Some authors \citep{oli00, bha00,sta04,cha09b} have considered possible different mechanisms where protons up to ultra high energies can be accelerated. Thus, we suppose that Cen A is capable of accelerating protons up to ultra high energies with a power law injection spectrum \citep{gup08},

\be\label{spepr}
\frac{dN_p}{dE_p}=A_p\,E_p^{-\alpha_p}
\ee

\noindent where $\alpha_p$ is the proton spectral index and $A_p$ is the proportionality constant.   Energetic protons in the jet mainly lose energy   by p$\gamma$  and pp interactions  \citep{ste68, ber90,bec09,ato03,cha09b}; as described  in the following subsections.

\subsection{p$\gamma$ interaction}

 The p$\gamma$; interaction takes place when accelerated protons collide with  target photons.  The single-pion production channels are $p+\gamma\to n+\pi^+$ and $p+\gamma\to p+ \pi^0$, where the relevant pion decay chains are $\pi^0\to 2\gamma$, $\pi^+\to \mu^++\nu_\mu\to e^++\nu_e+\bar{\nu}_\mu+\nu_\mu$ and $\pi^-\to \mu^-+\bar{\nu}_\mu\to e^-+\bar{\nu}_e+\nu_\mu+\bar{\nu}_\mu$ \citep{ato03}.

In this analysis we suppose  that protons interact with SSC photons ($\sim$ 150 keV) in the same knot. If so, the optical depth is given as $\tau_{p,ssc}\approx r_d\,\theta_{jet}\,\Gamma\, n^{obs}_{\gamma ssc} \sigma_{p\gamma}$, where  $r_d$ is the value of the dissipation radius \citep{bat03}, $\theta_{jet}$ is the jet aperture angle, $\sigma_{p\gamma}=0.9$ mbarn is the cross section  for the production of the delta-resonance in proton-photon interactions and  $n^{obs}_{\gamma ssc}$ is the particle density of SSC photons  into the observer frame \citep{bec09} given by,

\be
n^{obs}_{\gamma ssc}\approx \frac{\epsilon_{knot}\, L^{obs}}{4\pi\,r^2_d\,E^{obs}_{\gamma,c}}
\ee

Assuming that the luminosity of a knot along the jet is a fraction $\epsilon_{knot}\approx 0.1$ of the observed luminosity $L^{obs}=5\times 10^{43}\,erg\,s^{-1}$ for $E^{obs}_{\gamma,c}$ keV, the optical depth is,

\be
\tau_{p,ssc}\approx 8.2\times 10^{-7}\,\Gamma^{-1}\,\biggl(\frac{\theta_{\rm{jet}}}{0.3}\biggr)\,\biggl(\frac{\epsilon_{\rm{knot}}}{0.1}\biggr)\,\biggl( \frac{ L^{obs} } {5 \times10^{43}\,\rm{erg\, s^{-1}} } \biggr)\,\biggl( \frac{ r_d } {10^{16}\,\rm{cm} } \biggr)^{-1}\,\biggl( \frac{ E^{obs}_{\gamma,b}} {150\,\rm{keV} } \biggr)^{-1}\,.
\ee

The energy lost rate due to pion production is \citep{ste68, ber90},

\begin{equation}
t'_{p,\gamma}=\frac{1}{2\,\gamma_p}\int^\infty_{\epsilon_0}\,d\epsilon\,\delta_\pi(\epsilon)\xi(\epsilon)\,\epsilon\int^\infty_{\epsilon/2\gamma_p}dx\, x^{-2}\,n(x)
\end{equation}

\indent where $n(x)=dn_\gamma/d\epsilon_\gamma (\epsilon_\gamma=x)$, $\sigma_\pi(\epsilon)$ is the cross section of pion production for a photon with energy $\epsilon$ in the proton rest frame, $\xi(\epsilon)$ is the average fraction of energy transferred to the pion,  and $\epsilon_0=0.15$ is the threshold energy, $\gamma_p=\epsilon_p/m^2_p$.

The rate of energy loss,  $t'_{p,\gamma}$,   $f_{\pi^0,p \gamma}\approx t'_d/t'_{p,\gamma}$  (where $t'_d\sim r_d/\Gamma$ is the expansion time scale),  can  be calculated by following Waxman \& Bahcall 1997 formalism.

\begin{equation}
\label{}
f_{\pi^0, p \gamma}\approx \frac{(1+z)^2\,L^{obs}}{8\,\pi\,\Gamma^2\,\delta^2_D\,dt^{obs}\,E^{obs}_{\gamma,b}}\sigma_{\epsilon_{peak}}\,\xi({\epsilon_{peak}})\,\frac{\Delta\epsilon_{peak} }{\epsilon_{peak}}
\cases{
\frac{E^{obs}_p}{E^{obs}_{p,b}}       &  $E^{obs}_{p} < E^{obs}_{p,b}$ \cr
1                                                             &  $E^{obs}_{p} \geq E^{obs}_{p,b}$\cr
}
\end{equation}

Here, $\sigma_{peak} \approx 5\times\,10^{-28}$ cm$^2$ and $\xi({\epsilon_{peak}})\approx 0.2$ are the values of $\sigma$ and $\xi$ at $E_\gamma \approx \epsilon_{peak}$ and $\Delta\epsilon_{peak}  \approx 0.2$ GeV is the peak width.  

The differential spectrum, $dN_\gamma/dE_\gamma$ of the photon-pions produced by  p$\gamma$ interaction  is related to the fraction of 
energy lost through  the equation: $f_{\pi^0}(E_p)\,E_p\,dN_p/dE_p\,dE_p=E_\gamma\,dN_\gamma/dE_\gamma\,dE_\gamma $.  If we take into account  that $\pi^0$ typically carries $20\%$ of the proton's energy and that each produced photon shares the same energy then, we obtain the observed gamma  spectrum through  the  following relationship,

\begin{equation}
\label{pgamma}
\left(E^2\,\frac{dN}{dE}\right)_{\pi^0-\gamma} = A_{p,\gamma}
\cases{
\left(\frac{E_{\gamma}}{E_{0}}\right)^{-1} \left(\frac{E_{\gamma,c}}{E_{0}}\right)^{-\alpha_p+3}          &   $E_{\gamma} < E_{\gamma,c}$\cr
\left(\frac{E_{\gamma}}{E_{0}}\right)^{-\alpha_p+2}                                                                                        &   $E_{\gamma,c} < E_{\gamma}$\cr
}
\end{equation}

\noindent where
\be
A_{p,\gamma}=2.25\times 10^{-13}\frac{\delta_D^{\alpha_p} \,E_0^2\,A_p\,(11.1)^{2-\alpha_p}  e^{-\tau_{\gamma\gamma}}}{(1+z)^\alpha} \,\biggl(\frac{E^{obs}_{\gamma,c}}{150\,\rm{keV}}\biggr)^{-1}\,\biggl( \frac{ L^{obs} } {5 \times10^{43}\,\rm{erg\, s^{-1}} } \biggr)\,\biggl(\frac{dt^{obs}}{2.6 \times10^6 s}\biggr)\,\biggl(\frac{d_z}{3.8\,\rm{Mpc}}\biggr)^{-2}
\ee

\noindent and
\be
E^{obs}_{\pi^0-\gamma,c}=212.49\,\rm{GeV} \frac{\delta_D^2}{(1+z)^2} \biggl(\frac{E^{obs}_{\gamma,b}}{150\,\rm{keV}}\biggr)^{-1}
\ee

The eq.  \ref{pgamma} could represent the VHE photon contribution  to the spectrum.

\subsection{PP interaction}

Hardcastle et al. 2009 argues that the number density of thermal particles within the giants lobes is $n_p\sim 10^{-4}\,cm^{-3}$. If we assume that  the accelerated protons collide with this thermal particle target then, the energy lost rate due to pion production is given by \citep{ato03},

\begin{equation}
t'_{pp}=(n'_p\,k_{pp}\,\sigma_{pp})^{-1}
\end{equation}

\noindent where $\sigma_{pp}=30$ mbarn is the nuclear interaction cross section, $k_{pp}=0.5$ is the inelasticity coeficient and $n'_p$ is the comoving thermal particle density.  The fraction of energy lost by pp is  $f_{\pi^0,pp}\approx t'_d/t'_{pp}$ then,

\begin{equation}
f_{\pi^0 ,pp}=R\,n'_p\,k_{pp}\,\sigma_{pp}
\end{equation}

\noindent where R is the distance to the lobes from the AGN core.

The differential spectrum, $dN_\gamma/dE_\gamma$ of the photon-pions produced by  pp interaction  is related to the fraction of energy lost through the
equation:  $f_{\pi^0 , pp}(E_p)\,E_p\,(\frac{dN_p}{dE_p})\,dE_p=E_\gamma\,(\frac{dN_\gamma}{dE_\gamma})\,dE_\gamma$. Taking into account that  photon carries 18$\%$ of the proton energy,  we have that the observed pp spectrum is given by \citep{gup08},

\begin{equation}
\label{pp}
\left(E^{2}\, \frac{dN}{dE}\right)_{pp,\gamma}= A_{pp}\, \left(\frac{E_{\gamma}}{E_{0}}\right)^{2-\alpha_p}
\end{equation}
where,

\be
A_{pp}=9.97\times 10^{-21}  \frac{\Gamma^2\,\delta_D^2\,E_0^2\,A_p\,e^{-\tau_{\gamma\gamma}}}  {(1+z)^2}\biggl(\frac{R}{100\, \rm{kpc}}\biggr)\,\biggl(\frac{n'_p}{10^{-4}\,\rm{cm^{-3}}}\biggr)\,\biggl(\frac{dt^{obs}}{2.6 \times10^6 s}\biggr)^2\,\biggl(\frac{d_z}{3.8\,\rm{Mpc}}\biggr)^{-2}
\ee

The eq.  \ref{pp} could represent the VHE photon contribution  to the spectrum. 


\section{Calculation of physical parameters and expected UHECRs}

\begin{figure}
\plotone{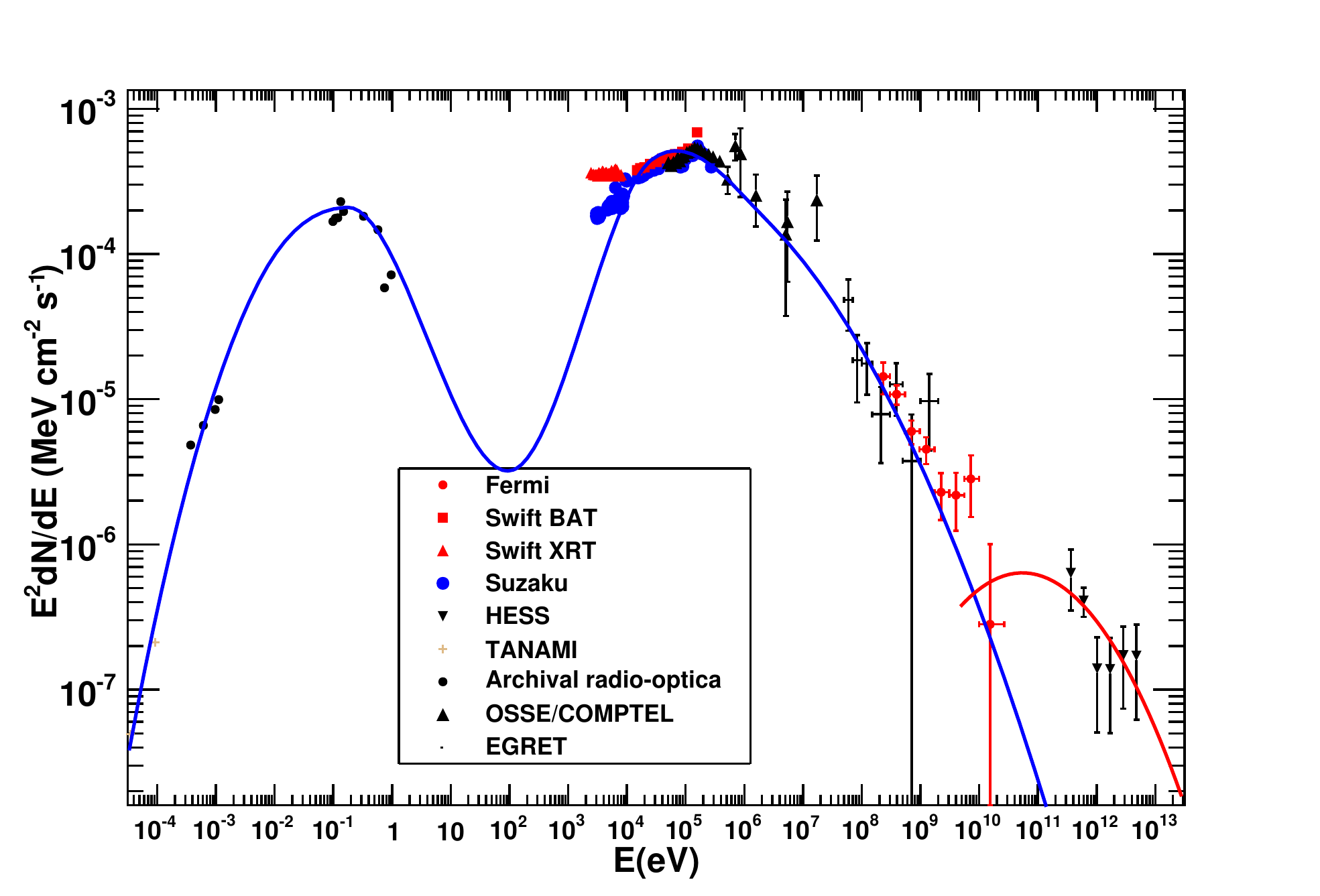}
 \caption{Fit of observed spectral energy distribution (SED) of Cen A.  The blue line is a fit to the broadband SED using the leptonic model described in section 2, while the red curve is the p$\gamma$ contribution described in section 3.}
 \label{SEDpgama}
 \end{figure}

\begin{figure}
 \plotone{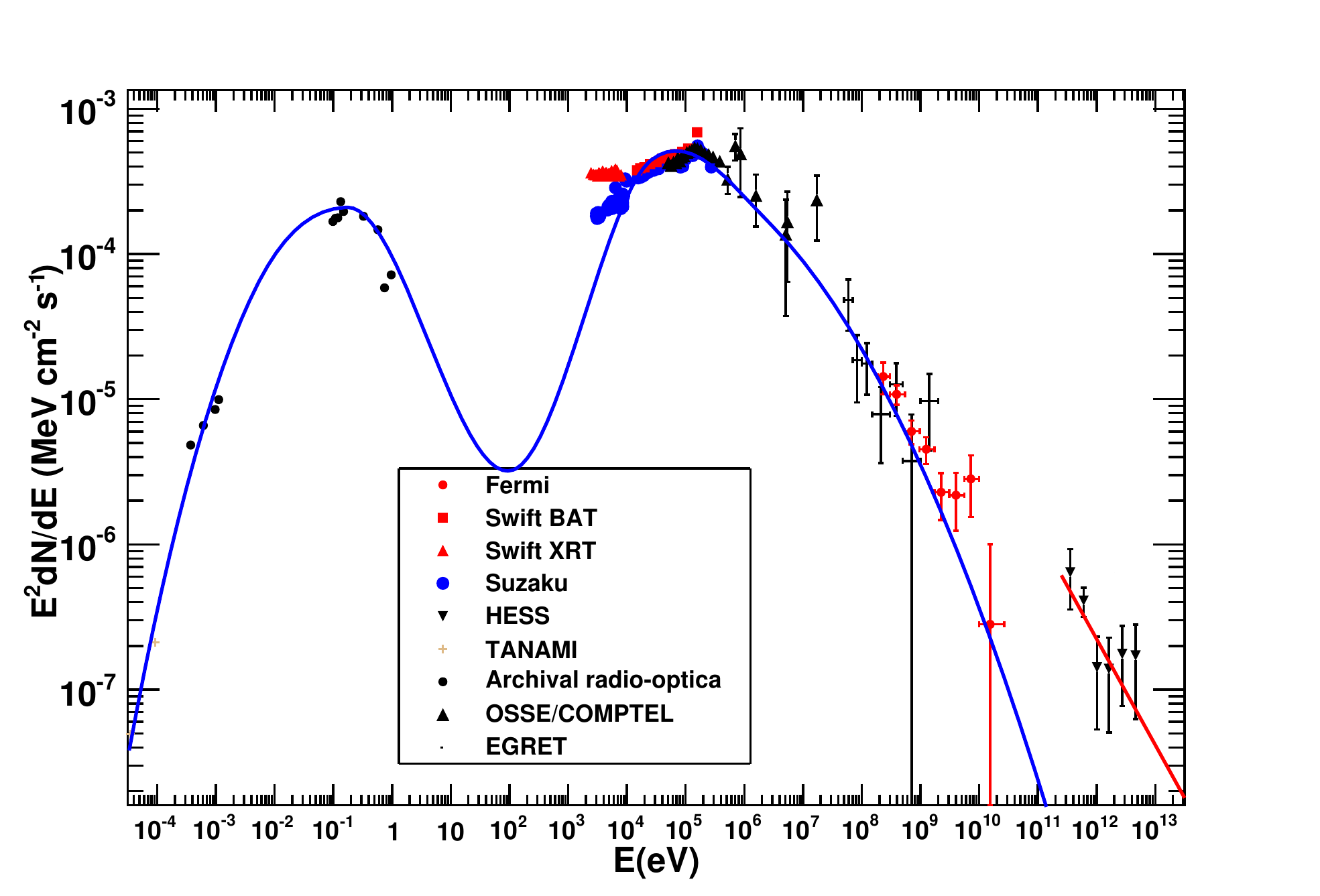}
    \caption{Fit of observed spectral energy distribution (SED)
    of Cen A.  The blue line is a fit to the broadband SED using the leptonic model described in section 2, while the red curve is the pp contribution described in section 3.}
    \label{SEDpp}
\end{figure}

A broadband fit to the SED of Cen A (data from Abdo et al. 2010) using our leptonic model (blue line) plus either p$\gamma$ or pp emission is shown in Figures \ref{SEDpgama} and \ref{SEDpp} respectively.   For this fit, we have adopted typical values reported in the literature such as luminosity ($L^{obs}$), variability ($dt^{obs}$),  thermal particle target density ($n_p$) and lobes distance ($R$)\citep{abd10, der09,  har09, rom96}. The viewing angle was chosen in accordance with the observed infrared range \citep[see, e.g.] [and reference therein]{hor06}. Then, from the fit we obtain the values of the  bulk Lorentz factor ($\Gamma$),  ratio of expansion time ($f_{es}$),    proportionality constants    ($A_e$, $A_{ic}$, $A_{p\gamma}/A_{pp}$),  magnetic field parameter  ($\xi_B$), electron parameters  ($\xi_e$) and  spectral index ($\alpha$).  Other quantities as magnetic field ($B$), electrons Lorentz factors ($\gamma_{e,min}$, $\gamma_{e,c}$ ), comoving radius ($r_d$), etc, are deduced these parameters. Table 1 shows all the values for used, obtained and deduced parameters in and from the fit.

\noindent The fit of  the VHE photon spectrum with a hadronic model (either p$\gamma$ or pp interaction) determines the spectral index $\alpha_p$,  energy normalization $E_0$ and proportionality constant $A_p$ (see section 2). So, we calculate the number of  UHECRs expected on Earth. Results are given in Table 1. As shown, the expected number of UHECRs  is extremely high if we consider  the VHE spectral gamma contribution to come from  p$\gamma$ interactions, while considering pp interactions the expected number of UHECRs is in agreement with PAO observations.

\begin{center}\renewcommand{\arraystretch}{0.75}\addtolength{\tabcolsep}{-1pt}
\begin{tabular}{ l c c }
  \hline \hline
 \large{Name} & \large{Symbol} & \large{Value} \\
 \hline

\hline
\hline
\normalsize{Input parameters to the model} & &  \\
\hline
\hline
\scriptsize{Variability timescales (s)} & \scriptsize{$dt^{obs}$} & \scriptsize{$2.5 \times 10^{6}\,$} \tiny {\citep{abd10}} \\
\scriptsize{Luminosity (erg s$^{-1}$)} & \scriptsize{$L^{obs}$} & \scriptsize{$5 \times 10^{43}\,$} \tiny {\citep{abd10}} \\ 
\scriptsize{Jet angle} (degrees)& \scriptsize{$\theta$} & \scriptsize{$40\,$} \tiny {\citep{hor06}} \\
\scriptsize{Normalization constant (leptonic process) } (MeV)& \scriptsize{$E_{0}$} & \scriptsize{0.1\,}\tiny {\citep{jou93}}\\ 
\scriptsize{Normalization constant  (p$\gamma$ process)} (TeV) & \scriptsize{$E_{0}$} & \scriptsize{1\,} \tiny {\citep{aha09}} \\ 
\scriptsize{Normalization constant (pp process) } (TeV)& \scriptsize{$E_{0}$} & \scriptsize{1\,} \tiny {\citep{aha09}} \\
\scriptsize{Thermal particle target density in lobes (cm$^{-3}$)} & \scriptsize{$n_{p}$} & \scriptsize{$1 \times 10^{4}\,$} \tiny {\citep{har09}} \\
\scriptsize{Lobes distance} (kpc) & \scriptsize{$R$} & \scriptsize{100\,} \tiny {\citep{har09}} \\

\hline
\hline
\normalsize{Calculated parameters with the model} & &  \\
\hline
\hline

\scriptsize{Bulk Lorentz factor} & \scriptsize{$\Gamma$} & \scriptsize{2.06 $\pm$ 0.03}\\
\scriptsize{Electron spectral index} & \scriptsize{$\alpha$} & \scriptsize{2.837 $\pm$ 0.004 } \\
\scriptsize{Magnetic field parameter}  & \scriptsize{$\epsilon_B$}  &  \scriptsize{0.1073 $\pm$ 0.0008 }\\
\scriptsize{Electron  energy parameter}  & \scriptsize{$\epsilon_e$}  &  \scriptsize{0.79 $\pm$ 0.14 }\\
\scriptsize{Ratio of expansion time}  & \scriptsize{$f_{es}$}  &  \scriptsize{0.0385 $\pm$ 0.0003}\\
\scriptsize{Proportionality electron constant} $(eV cm^2 s)^{-1}$  & \scriptsize{$ A_e $}  &  \scriptsize{ $(4.368\pm 0.003)\times 10^{15}$}\\
\scriptsize{Proportionality IC constant} $(eV cm^2 s)^{-1}$  & \scriptsize{$ A_{ic} $}  &  \scriptsize{ $(9.65\pm 0.07)\times 10^{16}$}\\
\scriptsize{Proportionality proton constant} $(TeV cm^2 s)^{-1}$  & \scriptsize{$ A_{pp} $}  &  \scriptsize{ $(5.9\pm 0.4)\times 10^{-7}$}\\
\scriptsize{Proportionality proton constant} $(TeV cm^2 s)^{-1}$  & \scriptsize{$ A_{p\gamma} $}  &  \scriptsize{ $(1.37\pm 0.99)\times 10^{4}$}\\
\scriptsize{Proton spectral index} & \scriptsize{$\alpha_p$} & \scriptsize{2.805 $\pm$ 0.008 } \\

\hline
\hline
\normalsize{Derived quantitatives} & &  \\
\hline
\hline

 \scriptsize{Doppler factor} & \scriptsize{$\delta_{d}$} & \scriptsize{1.47} \\
 \scriptsize{Magnetic field (G)} & \scriptsize{$B$} & \scriptsize{0.19} \\
 \scriptsize{Comoving  radius (cm)} & \scriptsize{$r_d$} & \scriptsize{$3.8 \times 10^{16}$} \\
\scriptsize{Minimum electron Lorentz factor} & \scriptsize{$\gamma_{m}$} & \scriptsize{$1.36 \times 10^3$} \\
 \scriptsize{Break electron Lorentz factor} & \scriptsize{$\gamma_{c}$} & \scriptsize{$3.47 \times 10^3$} \\
 
 \scriptsize{Apparent  UHECR Luminosity (p$\gamma$)}  ($erg\,s^{-1}$)& \scriptsize{$L_p$} & \scriptsize{$2.7\times 10^{49}$ } \\
 \scriptsize{Apparent  UHECR Luminosity (pp)}  ($erg\,s^{-1}$)& \scriptsize{$L_p$} & \scriptsize{$2.9\times 10^{39}$ } \\

\hline
\scriptsize{Predicted number of events: p$\gamma$ interaction} & \scriptsize{$N_{ev,p \gamma}$} & \scriptsize{$8.371\times 10^{10}$ } \\
\scriptsize{Predicted number of events: pp interaction} & \scriptsize{$N_{ev,p p}$} & \scriptsize{2.29} \\

\hline

 \end{tabular}
\end{center}

\begin{center}
\scriptsize{\textbf{Table 1. Parameters used and obtained  from and in the fit of  the spectrum of Centaurus A.}}\\
\scriptsize{}
\end{center}

\section{Neutrino expectation for Cen A }

The principal neutrino emission processes in the AGNs are hadronic. These interactions produce both, high energy neutrinos and high energy gamma rays, through pionic decay. As we mentioned before,  hadronic interactions generate mainly  pions by $p+p\rightarrow\pi^{0}+\pi^++\pi^-+ X$  (where $X$ is an hadronic product) and   $p+\gamma\rightarrow\Delta^{+}\rightarrow\pi^{0}+\pi^{+}$. The resulting neutral pion decays into two gamma rays, $\pi_{0}\rightarrow\gamma\gamma$, and the charged pion into leptons and neutrinos, $\pi^{\pm}\rightarrow e^{\pm}+\nu_{\mu}/\bar{\nu}_{\mu}+\bar{\nu}_{\mu}/\nu_{\mu}+\nu_{e}/\bar{\nu}_{e}$. The effect of neutrino oscillations on the expected flux balances the number of neutrinos per flavor \citep{bec08} arriving at Earth. Therefore, the measured emission of high energy gamma rays from AGNs suggests the possibility to have an equivalent high-energy neutrino flux. In the case of Cen A the redshift is $z=0.0018$, therefore  VHE  photons  are not  absorbed from EBL and we can consider the observed high energy gamma ray spectra as the intrinsic spectra emitted by this source and we can use it for the neutrino flux estimation.

Concerning the physics environment of Cen A, the optimistic conditions assumed to calculate the neutrino expectations are the following,
\begin{enumerate}
  \item The high energy gamma ray flux detected by HESS are produced according to the pp hadronic scenario in Cen A.
  \item The considered neutrino flux correlated to high energy gamma ray activity has a minimum duration of 1 year (i.e. the source is assumed to be stable).
  \item The neutrino spectrum of Cen A is assumed without any cut-off.
  \item The observed gamma-ray spectrum is considered as the intrinsic spectrum of Cen A.
\end{enumerate}

Considering that neutrinos and gamma rays are produced by the same hadronic interaction (pp), we follow the description of Becker(2008) to correlate these two messengers and we assume the neutrino spectrum to be the same as the VHE gamma spectrum recorded by H.E.S.S. Therefore we perform a Monte Carlo simulation of a possible Km$^3$ neutrino telescope in the Mediterranean sea in order to calculate the expected neutrino event rate. We choose this location  to have a good sensitivity with respect to the position of Cen A.

The Monte Carlo simulation takes into account the neutrino source position, the propagation of neutrino through water, the charged current  interaction with the respective muon production, the Cherenkov light produced by the muon, the photons produced by the electromagnetic showers
and the response of the simulated neutrino telescope. Then we calculate the signal to noise ratio in the telescope. 

\begin{figure}
\plotone{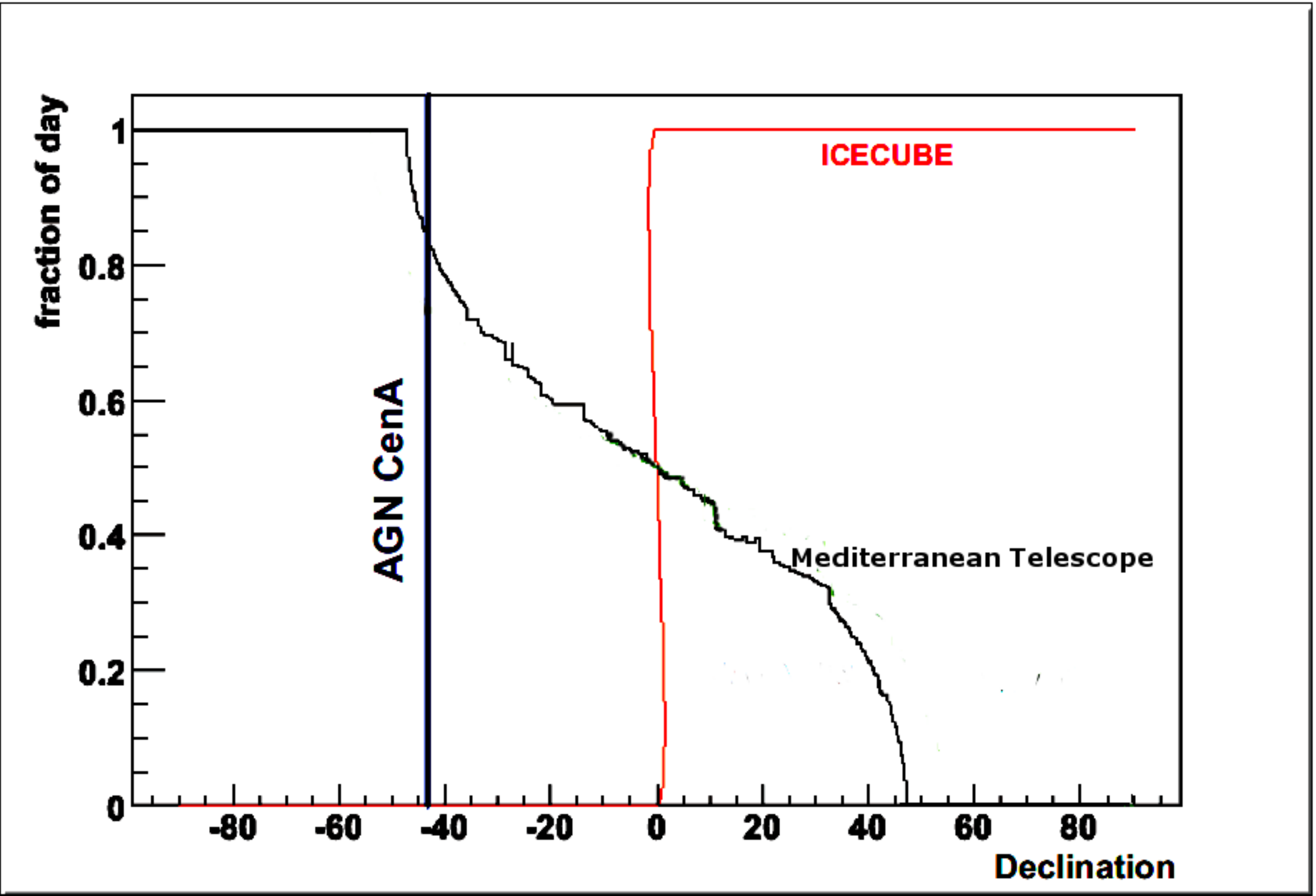}
  \caption{Calculated the Fraction of Day (FoD) when a source with a given declination is below the horizon with respect to our hypothetical Km$^{3}$ Telescope (solid line) in the Mediterranean and ICECUBE experiments (dashed line). For  Cen A, FoD is $\approx 0.8$ when Km$^{3}$ Telescope is considered. For ICECUBE, Cen A is all the time above the horizon.}
  \label{fod}
\end{figure}

In this analysis the neutrino ``backgrounds'' are represented by atmospheric neutrinos and cosmic diffuse neutrinos. The atmospheric neutrino ``background'' is generated by the interaction of high energy cosmic rays with nuclei in the atmosphere. The cosmic diffuse neutrino  ``background'' is taken as the average rate of neutrinos generated by all the galactic and extragalactic non-resolved sources. This cosmic diffuse neutrino flux  is discussed by Waxman and Bahcall  \citep{bah01, wax98} and his upper limit is given as $E^{2}_{\nu}\,d\Phi/dE_{\nu}<4.5\times 10^{-8}GeVcm^{-2}s^{-1}sr^{-1}$. The atmospheric neutrino flux implemented in our Monte Carlo is well described by the Bartol model \citep{bar04,bar06}  in the range between 10 GeV and 100 TeV.  We do not consider the ``background'' from atmospheric muon flux since it is filtered out by the Earth because Cen A is most of the time under the horizon for our hypothetical telescope,  see  Fig.\ref{fod}. For the calculation of signal to noise ratio we take  into account only the ``background''  inside the portion of the sky covered by a cone centered in the Cen A position and having an opening angle of 1$^\circ$ . This selection is motivated by the angular resolution of our neutrino telescope.  

\begin{figure}
\plotone{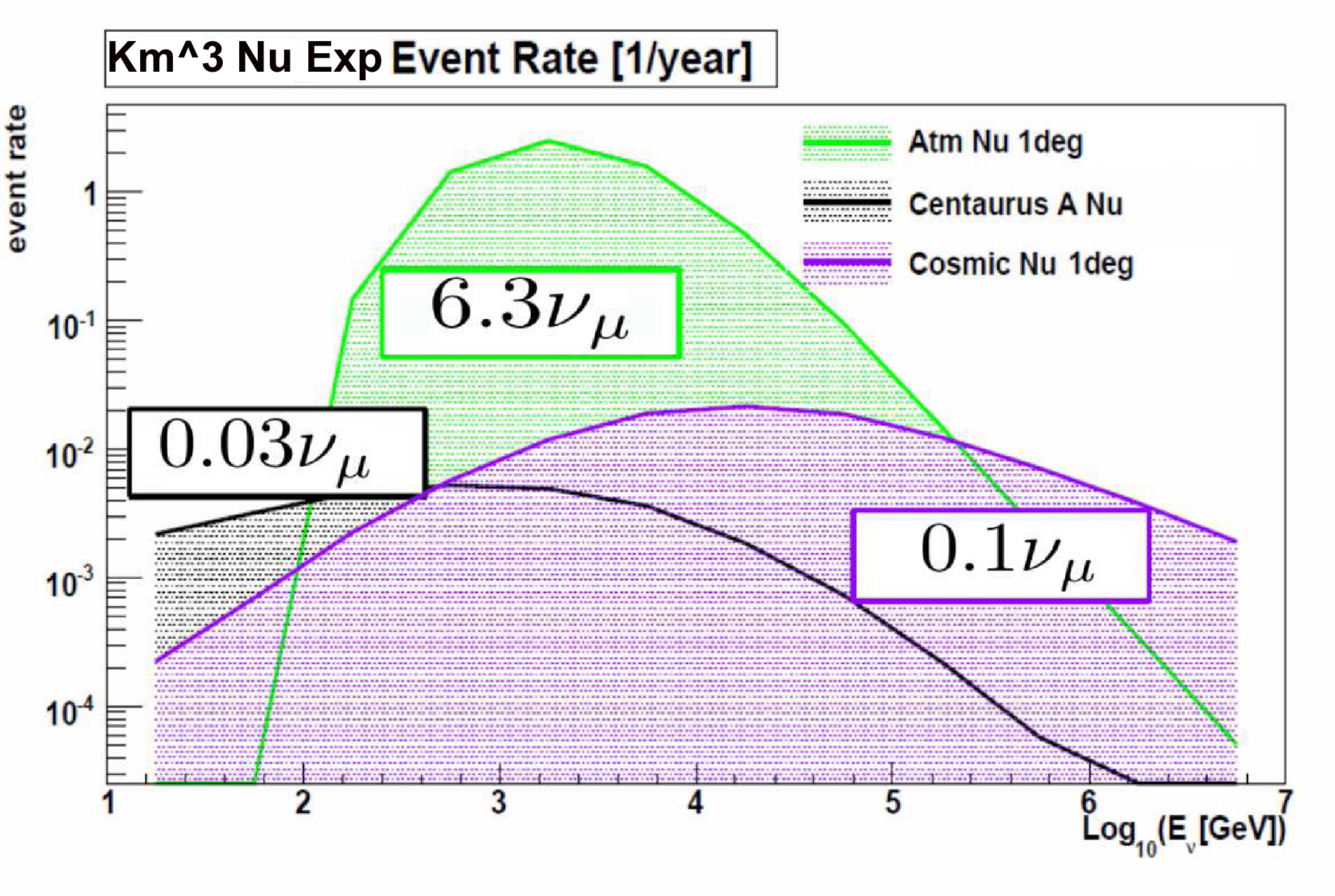}
   \caption{Neutrino event rate as a function of neutrino energy. The expected Centaurus A event rate for one year of observation with a $Km^{3}$ neutrino telescope assuming the spectrum measured by HESS (black line), the event rate for atmospheric neutrinos (green line) and cosmic neutrinos (purple line) considering the solid angle of $1^{\circ}$ around the sources. A quality cut has been applied to reconstructed events. The integrated number of neutrinos is given for each case.}
   \label{CenAevtrateHessBartWax}
\end{figure}

Using  the assumed neutrino spectrum we obtain for the Km$^3$ telescope the expected neutrino event rate shown in Fig. \ref{CenAevtrateHessBartWax}.  As  observed, the integrated signal neutrino event rate in one year of recording data is one order of magnitude below the cosmic neutrino event rate and two order of magnitude below the atmospheric neutrino event rate reconstructed in the region around Cen A. Moreover, even considering few years of neutrino telescope operation, with the considered spectrum, we are not able to disentangle neutrino emission from Cen A.

\section{Summary and conclusions}

We have presented a leptonic and hadronic model to describe the broadband photon spectrum of Cen A.  Our model has eight free parameters   (equipartition magnetic field, equipartition electron energy,  bulk Lorentz factor, spectral  index, ratio of expansion time  and proportionality constants).
The leptonic model describes the spectrum up to a few GeV energies while the hadronic model describes the Cen A spectrum at TeV energies.
Two hadronic interactions have been considered, p$\gamma$ and pp interactions. In the first case, the target is considered as SSC photons with energy of $\sim 150$ keV, while in the second case, the target protons are those in the lobes of Cen A. Only one hadronic interaction is considered at the time but in both cases, the proton spectrum is extrapolated up to ultra high energies to estimate the number of UHECR events expected at Earth. We have required a good description of the photon spectrum to obtain values for the quantities required to estimate the UHECR events. 

When p$\gamma$ interaction is considered, the expected number of UHECR obtained is several orders of magnitude above the observed by PAO. However, when pp interaction is considered, the expected number of UHECR is in very good agreement with PAO observations. 

We have also calculated the neutrino event rate from pp interactions observed by a hypothetical Km$^{3}$ neutrino telescope in the Mediterranean sea. We have calculated the signal to noise ratio considering atmospheric and cosmic neutrino ``backgrounds''. We have obtained that the expected signal event rate is below the required one to disentangle the neutrino emission from Cen A from the "backgrounds".

\acknowledgments

We thank the anonymous referee for the comments given to improve the paper. We also thank to Charles Dermer, Markus B\"{o}tcher,  Parisa Roustazadeh,    Bin Zhang, Giulia DeBonis, Bachir Bouhadef, Mauro Morganti, Dario Grasso, Antonio Stamerra and Teresa Montaruli for useful discussions. 

This work was supported by DGAPA-UNAM (Mexico) Project Numbers IN112910 and IN105211 and Conacyt project number 105033.

\begin{thebibliography}{}

\bibitem[Abraham et al. (2007)]{abr07} Abraham J.  et al. (Pierre Auger Collaboration),  2007, Science, 318, 938
 \bibitem[Abraham et al. (2008)]{abr08}Abraham J.  et al. (Pierre Auger Collaboration),  2008, \apj, 29, 198.
\bibitem[Adbo et al. (2010)]{abd10} Abdo A. A. et al.    (FERMI Collaboration),  2010, \apj, 719, 1433.
\bibitem[Aharonian et al. (2005)]{aha05} Aharonian F.  et al.  (HESS Collaboration),  2005, \aap, 441, 465.
\bibitem[Aharonian et al. (2009)]{aha09} Aharonian F.  et al.  (HESS Collaboration),  2009, \apj, 695, L40.
\bibitem[Allen (1993)]{all93}Allen  W.H.  et al.  (JANZOS Collaboration), 1993, \apj, 405, 554.
\bibitem[Atoyan \& Dermer  (2003)]{ato03}Atoyan A. M \& Dermer C. D., 2003, \apj, 589, 79.
\bibitem[Baity et al. (1981)]{bai81} Baity  W. A. et al., 1981, \apj 244, 429.
\bibitem[Bahcall \& Waxman (2001)]{bah01}Bahcall  J. \& Waxman E., 2001, Physical Review D., 64(2): 023002.
\bibitem[Barr  et al. (2004)] {bar04} Barr G.D., Gaisser T. K. , Lipari  P., Robbins S.  \&  Stanev T., Phys. Rev. D 70, 023006 (2004).
\bibitem[Barr  et al. (2006)] {bar06} Barr G.D., Gaisser T. K., Robbins S., \& Stanev T., Phys. Rev. D 74, 094009 (2006).
\bibitem[Becker (2008)]{bec08}Becker J. K., 2008, Physics Reports, 458 , 173.
\bibitem[Becker \& Biermann (2009)]{bec09}Becker J. K. \& Biermann P. L., 2009, Astropart. Phys. 31, 138.
\bibitem[Berezinskii et al. (1990)]{ber90} Berezinskii, V.S., Bulanov S. V., Dogiel V. A., Ginzburg V. I.  \& Ptuskin V. S. 1990, Astrophysics pf Cosmic Rays (North-Holland:Amsterdam), Ch. 4.%


\bibitem[Bhattacharjee \&  Gupta (2003)]{bat03}Bhattacharjee P. \&  Gupta N., 2003, Astropart. Phys., 20, 169.%
\bibitem[Bhattacharjee \&  Sigl (1998)]{bha00}Bhattacharjee P. \&  Sigl G., 2000, Phys. Rept., 327, 109.%
\bibitem[Blandford \&  McKee (1976)]{bla76}Blandford R. D. \&  McKee C. F., 1976,  Phys. Fluids. 19,  1130.%
\bibitem[Bloom  \& Marscher (1996)] {blo96}  Bloom, S. D.  \& Marscher, A. P. 1996, \apj, 461, 657.
\bibitem[Bowyer et al. (1970)]{bow70}Bowyer  C. S., Lampton M., Mack J.  \&  De Mendonca F., 1970, \apj 161 L1.
\bibitem[Carrami\~nana et al. (1990)]{car90}Carrami\~nana et al.,  1990, \aap, 228, 327.
\bibitem[Cheng \& Wei (1996)]{che96}Cheng K. S. \& Wei D. M.  1996,  \mnras, 283, L133.%
\bibitem[Chiaberge et al.( 2001)]{chi01} Chiaberge M., Capetti A. \& Celotti  2001, \mnras, 324, L33.
\bibitem[Cuoco \& Hannestad (2008)]{cuo09}Cuoco A.  \& Hannestad S., 2008, Phys. Rev. D, 78, 023007.
\bibitem[Dermer  et al. (2009)]{der09}Dermer C. D., Razzaque S., Finke J. D.  \& Atoyan A., 2009, New J. Phys. 11,  065016.
\bibitem[Dermer \& Menon (2009)]{cha09b}Dermer C. D. \& Menon G. 2009,  High energy radiation from  Black Holes, Princeton University Press. 2009.%
\bibitem[Engel et al. (2007)]{eng07}Engel R.l R  (The Pierre Auger Collaboration), 2007, arXiv:  0706.1921.
\bibitem[Fragile et al. (2004)]{fra04}Fragile P. C., Mathews G., Poirier J.  \&  Totani T., 2004, Astropart. Phys.  20, 591.
\bibitem[Gallant  (1999)]{gal99}Gallant Y. A., 2002, Lectures Notes in Physics,  589,  24.%
\bibitem[Gopal-Krishna et al. (2010)]{gop10}  Gopal-Krishna, Biermann P.  L. , De Souza V.  \&  Wiitta P. J.,  2010, Astropart. Phys.  720, L155.%
\bibitem[Gorbunov  et al. (2008)]{gor08} Gorbunov D., Tinyakov, P. Tkachev, I \& Troitsky, S., 2008, Sov. J. Exp. Theor. Phys. Lett, 87, 461.
\bibitem[Grindlay et al. (1975)]{gri75}Grindlay J. E, Helmken H. F., Brown R. H.,Davis, J. \& Allen L. R.,  1975, \apj, 197, L9.
\bibitem[Gupta \& Zhang (2007)]{gup07}Gupta N. \& Zhang B.   2007,  \mnras, 380, 78.%
\bibitem[Gupta (2008)]{gup08}Gupta N., 2008, JCAP, 7060806, 022.
\bibitem[Hague et al. (2009)]{pa2009}Hague J. D.,(Pierre Auger Collaboration), 2009, Proc. 31st ICRC, Lodz
\bibitem[Hardcastle et al. (2003)] {har03}Hardcastle  M. J., Worrall  D. M., Kraft  R. P.,  Forman W. R., Jones  C.  \& Murray S. S., 2003,  \apj, 593, 169.
\bibitem[Hardcastle  et al. (2009)]{har09} Hardcastle M. J., Cheung C. C., Feain I. J. \& Stawarz L., 2009, \mnras, 393, 1041.
\bibitem[Hinshaw et al. (2009)]{hin09}Hinshaw et al., 2009, \apjs, 180, 225.
\bibitem[Honda (2009)]{hon09}Honda M., 2009, \apj, 706, 1517.
\bibitem[Horiuchi et al. (2006)]{hor06}Horiuchi S., Meier D. L., Preston R. A. \& Tingay S. J.,  2006, PASJ, 58, 211.
\bibitem[Isola  et al. (2002)]{iso02}Isola, C., Lemoine M. \& Sigl G., 2002, Phys. Rev. D, 65, 023004.
\bibitem[Jourdain  et al. (1993)]{jou93}Junkes  E. et al.,    1993, \apj, 412, 586.
\bibitem[Junkes et al.(1993)]{jun93}Junkes  N., Haynes, R. F., Harnett, J.I., \& Jauncy, D. L. 1993, \aap, 269,29.
\bibitem[Kabuki et al. (2007)]{kab07}Kabuki, S. et al. (CANGAROO III Colaboration) 2007, \apj, 668, 968.
\bibitem[Kachelriess  et al. (2009a)]{kac09a} Kachelriess M., Ostapchenko S. \& Tomas R., 2009, New J. Phys. 11,  065017.
\bibitem[Kachelriess  et al. (2009b)]{kac09b} Kachelriess M., Ostapchenko S. \& Tomas R., 2009, Int. J. Mod. Phys. D 18, 1591.
\bibitem[Lenain et al. (2008)]{len08}Lenain J. P.,  Boisson C.,  Sol H.  \& Katarzy«nski  K.,  2008  \aap 478, 111.
\bibitem[Moskalenko  et al. (2009)]{mos09}Moskalenko, I. V., Stawarz, L., Porter T. A. \&  Cheung, C. C., 2009, \apj, 693,  1261.
\bibitem[Mushotzky \& Baity (1976)]{mus76}Mushotzky R. F.,   Baity W. A.,   Wheaton W.A., \&  Peterson L. E., 1976 , \apj, 206, L45.
\bibitem[Olinto (2000)]{oli00}Olinto A. V., 2000,  Phys. Rept. 333, 329.%
\bibitem[Orellana \&  Romero (2009)]{ore09} Orellana M.  \&  Romero G. E.,  2009   AIP Conf.  Proc., 1123,  242.
\bibitem[Raue  \&  Mazin (2008)]{rau08}Raue  M. \&  Mazin D. ,  2008,  International Journal of Modern Physics 17, 1515.
\bibitem[ Rieger  \&   Aharonian   (2009)]{rie09} Rieger F. M.  \&   Aharonian F. A., 2009, \aap, 506, L41.%
\bibitem[Romero  et al. (1996)]{rom96}Romero G. E., Combi J. A., Anchordoqui L.A. \& Perez S. E., 2009, Astropart. Phys., 5, 279.
\bibitem[Rowell et al. (1999)]{row99}Rowell, G. P. et al. (CANGAROO III Colaboration) 1999, Astropart. Phys. 11, 217.
\bibitem[Rybicki \& Lightman (1979)]{ryb79}Rybicki G.B. \& Lightman A.P.1979,  Radiative processes in Astrophysics (Wiley, New York. 1979).%
\bibitem[Sreekumar et al. (1999)]{sre99}Sreekumar  P., Bertsch D. L., Hartman R. C., Nolan P.L. \& Thompson D. J., 1999, \apj, 11, 221.
\bibitem[Stanev (2004)]{sta04}Stanev T., 2004,  High Energy Cosmic Rays, Springer,  2004.%
\bibitem[Stecker  (1968)]{ste68}Stecker F. W.   1968,  Phys. Rev. Lett., 21, 1016.%
\bibitem[Steinle et al. (1998)]{ste98}Steinle H. et al.,  1998, \aap, 330, 97.
\bibitem[Tavecchio  et al. (1998)]{tav98} Tavecchio, F., Maraschi, L.  \&  Ghisellini, G., 1998, \apj, 509,  608.
\bibitem[Unger  et al. (2007)]{ung07}Unger M., Engel R., Schussler F., Ulrich R. \& Pierre Auger Collaboration,   2007,  Astron. Nachrichten, 328, 614.
\bibitem[Unger  et al. (2008)]{ung08}Unger M., Dawson B.R., Engel R., Schussler F.  \& Ulrich R.,   2008,  Nucl. Instrum. Methods Phys. Res. A, 588, 433.
\bibitem[Vietri (1995)]{vie95}Vietri  M.,  1995 \apj, 453,  883.%
\bibitem[Waxman \& Bahcall (1997)]{wax97}Waxman E. \& Bahcall J., 1997,  Phys. Rev. Lett. 78, 12.
\bibitem[Waxman \& Bahcall (1998)]{wax98}Waxman E. \& Bahcall J., 1998,  Texas simposium on Relativistic Astrophysics and Cosmology, December 1998. 
\bibitem[Winkler et al. (1975)]{win75}Winkler  F. P. \&  White A. E. 1975 \apj, 199, L139.
\bibitem[Yamamoto  et al. (2007)]{yam07}Yamamoto  T. \& Pierre Auger Collaboration,   2007,  arXiv:0707.2638.

\end{thebibliography}
\end{document}